\documentclass[pre,showpacs, showkeys]{revtex4}
\usepackage{graphicx}
\usepackage{amssymb}

\newcommand{\beq}[0]{\begin{equation}}
\newcommand{\eeq}[0]{\end{equation}}
\newcommand{\e}{\epsilon}
\newcommand{\thet}{\vartheta}
\newcommand{\la}{\langle}
\newcommand{\ra}{\rangle}
\newcommand{\ds}{\displaystyle}
\newcommand{\avh}{\la H_1 \ra}
\newcommand{\pa}{\partial}
\newcommand{\w}{\omega}
\newcommand{\ud}{{\mathrm d}}

\begin{document}

\title{Existence of multi-site intrinsic localized modes in one-dimensional Debye crystals}

\author{
V. Koukouloyannis$^{1, }$\footnote{Electronic address:
\texttt{vkouk@physics.auth.gr} ;
\texttt{http://users.auth.gr/$\sim$vkouk} .}, I. Kourakis$^{2,
}$\footnote{Research partly carried out at: Universiteit Gent,
Sterrenkundig Observatorium, Krijgslaan 281, B-9000 Gent, Belgium;
\\ Electronic address: \texttt{ioannis@tp4.rub.de} ;
\texttt{www.tp4.rub.de/$\sim$ioannis} .}} \affiliation{$^{1}$
School of Physics, Theoretical Mechanics, Aristotle University of Thessaloniki
\\54124 Thessaloniki, Greece\\
$^{2}$ Institut f\"ur Theoretische Physik IV, Fakult\"at f\"ur
Physik und Astronomie, Ruhr Universit\"at Bochum, D-44780 Bochum,
Germany}
\date{\today; last revised by IK}

\begin{abstract}
The existence of highly localized multi-site oscillatory
structures (discrete multibreathers) in a nonlinear Klein-Gordon
chain which is characterized by an inverse dispersion law is
proven and their linear stability is investigated. The results are applied in the description of vertical (transverse,
off-plane) dust grain motion in dusty plasma crystals, by taking
into account the lattice discreteness and the sheath electric
and/or magnetic field nonlinearity. Explicit values from
experimental plasma discharge experiments are considered. The
possibility for the occurrence of  multibreathers associated with
vertical charged dust grain motion in strongly-coupled dusty
plasmas (dust crystals) is thus established. From a fundamental point of view, this study aims at providing a first
rigorous investigation of the existence of intrinsic localized modes in Debye crystals and/or dusty
plasma crystals and, in fact, suggesting those lattices as model systems
for the study of fundamental crystal properties.

\end{abstract}

\maketitle

\section{Introduction}

Periodic lattices of interacting particles are known from solid
state physics to sustain, apart from propagating vibrations
(\textit{phonons}), a variety of localized excitations, due to a
mutual balance between the intrinsic nonlinearity of the medium
and mode dispersion. Such structures, ``traditionally'' sought for
and investigated in a continuum approximation (i.e. assuming that
the typical spatial variation scale far exceeds the typical
lattice scale, e.g. the lattice constant $r_0$), include
non-topological \emph{solitons} (pulses), \emph{kinks} (i.e.
shocks or dislocations) and localized modulated envelope
structures (\emph{envelope solitons}). Various generic nonlinear
theories have been developed in order to investigate their
occurrence in different physical contexts \cite{DauxoisP}. In
addition to these (continuum) theories, which neglect discreteness
for the sake of analytical tractability, attention has been paid
since more than a decade ago to highly localized vibrating
structures [\emph{discrete breathers} (DBs) or \textit{intrinsic
localized modes} (ILMs)], which owe their existence to the lattice
discreteness. Following some pioneering ILM related works in the
late 80's (see e.g., in Refs. \cite{Page, Takeno, Dauxois,
Kivshar, Campbell1}, the breakthrough in the theoretical study of
DBs took place with the appearance of the first rigorous proofs of
discrete breather existence, furnished independently by R.S.
MacKay and S. Aubry \cite{MacKay1} (who introduced the notion of
continuation from a suitable anticontinuous limit) and S. Flach
\cite{Flach1} (using a homoclinic orbit approach). The former
method was later extended later in \cite{Sepulchre}. The discrete
vibrational state in which the energy of a localized excitation is
concentrated in more than one sites is called a {\it
multibreather}. An extensive study on the existence and stability
of multibreathers can be found in Refs.
\cite{ams,koukicht1,kouk,koukichtstability,arc}, while as regards the
homoclinic method approach on the same subject one is referred to
Refs.  \cite{bountis,bountis1}. A large number of noteworthy
detailed studies have appeared, elucidating various aspects
involved in the spontaneous formation, mobility and interaction of
DBs, both theoretically and experimentally; see e.g., in Refs.
\cite{Aubry,Flach2, MacKay2, Chaos, Campbell2} for a review.

Recent studies of collective processes in a dust-contaminated
plasma (DP) \cite{psbook, SVreview, SVbook} have revealed a variety 
of new linear
and nonlinear collective effects, which are observed in laboratory
and space dusty plasmas. An issue of particular importance in DP
research is the formation of strongly coupled DP crystals by
highly charged dust grains, typically in the sheath region above a
horizontal negatively biased electrode in experiments
\cite{psbook, Morfill}.
Low-frequency oscillations are known to occur in gas-discharge 
experiments \cite{Morfill} involving
these mesoscopic dust grain quasi-lattices, in the longitudinal
(in-plane, acoustic mode), horizontal transverse (in-plane) and
vertical transverse (off-plane) directions \cite{psbook, SVreview, SVbook, Morfill}.
Here, we shall only focus on the latter (transverse) degree of 
freedom, which was first predicted in Refs. \cite{VSC1, VSC2} (also see Ref. \cite{VSC3}) and 
later experimentally observed.

Nonlinearity is an intrinsic feature of dust crystal dynamics. It may 
be due either
to inter-grain (Debye-type, screened electrostatic) nonlinear
interactions, or to geometric effects, i.e., to coupling among 
different modes \cite{PhysScripta}.
Finally, and rather importantly, nonlinearity is also due to the 
plasma sheath environment, which provides the electric 
\cite{Morfill} and/or magnetic \cite{VMYaro} forces
ensuring equilibrium by suspending the crystal at a levitated 
position above the negative electrode;
thus, the electric/magnetic forces in the sheath provide a 
``substrate''' potential which is intrinsically anharmonic (see 
discussion below).
 From the point of view of nonlinear science, therefore,
DP crystal dynamics provide a blend of anharmonic on-site potential 
and coupling nonlinearity, accounting for a unique hybrid among
nonlinear Klein-Gordon type and, say, Fermi-Pasta-Ulam (FPU) type 
theories (cf. the discussion in \cite{IKIJBC}). Despite this fact,
present day knowledge of nonlinear mechanisms related to dust
lattice modes is admittedly still in a preliminary stage. Small
amplitude localized longitudinal excitations (described by a
\textit{Boussinesq} equation for the longitudinal grain
displacement $u$, or a \textit{Korteweg-deVries} equation for the
density $\partial u/\partial x$) were considered in Refs.
\cite{Melandso} and generalized in Ref. \cite{IKPKSEPJDsols}.
Also, the amplitude modulation of longitudinal \cite{AMS2,
IKPKSLDLWMI} and transverse (vertical, off-plane)
\cite{IKPKSTDLWMI, IKPKSTMDLWMI} dust lattice waves (LDLW, TDLW,
respectively) was recently considered. A model for nonlinear DP
lattice dynamics taking into account transverse-to-longitudinal
mode coupling was suggested in Ref. \cite{PhysScripta}. Those
studies were recently extended to diatomic-like Debye
(bi-)crystals \cite{IKPOPbi} and eventually extended to more
realistic, hexagonal crystalline geometries \cite{Farokhi}. All of
these studies have relied on a quasi-continuum description of dust
lattice dynamics. The \emph{discrete} character of dust-lattice
oscillations has, to our best knowledge, not yet been studied,
apart from a recent first investigation which was restricted to
single-mode transverse dust-breathers \cite{IKPKSPOPDB}. That
study has examined the properties of {\it vertical} (off-plane,
transverse) dust lattice vibrations, from first principles. A
description of such a system, using the homoclinic method
approach, was formulated in Ref. \cite{IKproc}, providing some
first results on breathers in Debye crystals. The existence of
breathers associated to {\it longitudinal} (in-plane, acoustic)
charged particle motion in a Debye crystal was discussed in Ref.
\cite{kourakis2}, although a detailed investigation from first
principles (which would be the analogue of the Fermi-Pasta-Ulam
problem, tailor-fit to Debye- or dust- crystals) is still missing

Most interestingly, the transverse (linear) dust lattice mode is
known to obey an \emph{inverse dispersion} law: therefore the
group velocity $v_g = \omega'(k)$ and the phase velocity $v_{ph} =
\omega/k$ point towards opposite directions. The \emph{anharmonic}
character of the vertical on-site potential (confirmed
experimentally \cite{Ivlev2000, Zafiu}), in combination with the
high \emph{discreteness} of dust crystals, clearly suggested by
experiments \cite{Liu, Misawa}, may play an important role in
mechanisms like energy localization, information storage and
response to external excitations. Although all of the ingredients
prescribing the occurrence of discrete breathers in DP crystals
seem to be present, rather surprisingly, this aspect of DP
dynamics still remains essentially unexplored.

From a fundamental point of view, crystalline arrangements of
charged particles (Debye crystals and/or dusty plasma lattices)
provide an excellent toy-model for the study of nonlinear
excitations in atomic-scale lattices, or in ionic scale
crystalline structures (e.g., ultra-cold plasmas, UCPs), since
they mimic atomic chain dynamics at a scale which is appropriate
for human observation (dust crystals can even be filmed and
analyzed via simple and inexpensive video image processing; see
e.g. in Ref. \cite{Morfill}). Here, we aim to stress that, as
regards Discrete Breather excitations, such lattices may provide a
convenient mesoscopic analogue of atomic chains, which should
allow for direct observation of localized excitations and
eventually in-situ confirmation of analytical theories, via
purpose-built experiments. To our best knowledge, no other meso-
or macroscopic system permitting such experimental modelling
freedom exists.

In this study, we are interested in investigating the conditions
for the occurrence of discrete multi-site lattice excitations
(\emph{multibreathers}) in a nonlinear (infinite sized)
Klein-Gordon-like chain, which is characterized by an inverse
dispersion law. Nonlinearity is supplied by a (non harmonic)
on-site potential, while inter-particle interactions are taken to
be linear. A negative coupling coefficient (``spring constant'')
value is assumed, in account of an inverse dispersion. Our results
are applied in a description of real transverse dust-lattice
excitations, as observed in plasma discharge experiments. The
formalism, the analytic tools and the numerical methods used in
the present study are described in
\cite{koukicht1,koukichtstability, kouk} and allow us to evaluate
the range of values of relevant physical parameters which may
account for breather formation.

\section{Existence and Stability of multi-site intrinsic localized modes}

Our aim is to prove the existence of multibreather excitations in
Debye crystals. The method we adopt is based on the continuation
of a specific state of a suitable anticontinuous limit, as e.g. in
\cite{MacKay1,Sepulchre}. The formalism used is described in Ref.
\cite{koukicht1}. A brief outline of the method is provided in the
following.

Consider the Hamiltonian describing a chain consisting of infinitely many identical nonlinear oscillators, with displacement $x_i$ and on-site potential $V(x_i)$, possessing  a stable equilibrium at $x_i=0$, i.e. ${V'}_i(0)=0$ and ${V''}_i(0)=\w_p^2>0$ 
(unit mass is assumed in the notation), coupled through a coupling constant $\e$
\beq H=H_0+\e H_1=\sum_{i=-\infty}^{\infty}\left(\frac{1}{2}p_i^2+V(x_i)\right)+\frac{\e}{2}\sum_{i=-\infty}^{\infty}(x_{i+1}-x_i)^2.
\label{hamfull} \eeq
 The corresponding equations of motion are
\beq\ddot{x}_i=-V'(x_i)+\e\,(x_{i+1}-2\,x_i+x_{i-1}) \qquad
\qquad \forall\, i\in \mathbb{Z} \, . \label{KG}
\eeq 
This is the
classical Klein-Gordon chain, which is well known to support
multibreather solutions. Note that, the multibreather existence
theorems, based on a continuation from a suitable anticontinuous
limit \cite{ams,koukicht1} hold for an $\e$-neighborhood around
zero, and are thus valid either for $\e>0$ or for $\e<0$, provided
that $|\e|<<1$.

Consider the integrable anticontinuous limit ($\e=0$) i.e. the
chain is consisting of uncoupled oscillators. In this limit we
consider the state where all the oscillators lie in equilibrium
apart from $n+1$ ``central'' ones which lie on periodic orbits
satisfying the resonance condition
\beq\frac{\w_0}{k_0}=\cdots=\frac{\w_n}{k_n}=\w \, , \qquad \qquad
k_i \in \mathbb{Z}.\label{rescon}\eeq This state is time--periodic
with period $T=2\pi/\w$ and trivially space--localized. We seek
the conditions under which this state will be continued for
$\e\neq0$ by keeping the previously mentioned attributes,
providing this way a multibreather. At this limit, the motion of
the central oscillators is described by
$$\begin{array}{rcll}
w_i&=&\w_it+\thet_i\\[8pt] J_i&=&\mathrm{const.}& \qquad \qquad
i=0, \ldots,  n \, ,
\end{array}$$
where $(w_i, J_i)$ are the action angle-variables of the uncoupled
oscillators, $\thet_i$ are the initial angles and $\w_i$ are the
corresponding angular frequencies. The $T$-periodic motion,
which is described by (\ref{rescon}), can be continued for $\e\neq0$
small enough, to form a $T$-periodic $(n+1)$-site breather,
provided that the following conditions hold: \\
1) The {\it anharmonicity condition} of the individual
oscillators, i.e. $\ds{\ud \w_i}/{\ud J_i}\neq0$, at least in the
neighbourhood
of the specific periodic orbit.\\
2) The {\it nonresonance condition}: $\w_{p}\neq m\,\w, \quad\forall
m\, \in\mathbb{N}$, where $\w_{p}$ denotes the linear frequency ot the single oscillator.
However, even if both of these conditions hold, not all the states
of the anticontinuous limit will be continued to a multibreather.
In addition, the phases of the oscillators in this limit must be
such that the system of equations \beq\frac{\pa \la H_1\ra}{\pa
\phi_i}=0\label{conz}\qquad \qquad \qquad i=1\ldots n \, \eeq has
simple zeros, i.e. it is also required that
$\det\left|{\partial^2\la H_1\ra}/{\partial \phi_i\pa
\phi_j}\right|\neq0$, where
$\phi_i=k_{i}\thet_{i-1}-k_{i-1}\thet_{i}$ is a
generalization of the notion of phase difference between the
successive oscillators, in order to include resonances other that
the $1:1$. Here, \beq\la H_1\ra=\int_0^TH_1\ud t\eeq is the
average value of the perturbative term of the Hamiltonian
calculated along a periodic orbit of the uncoupled system over a
time-period.

Since we consider identical oscillators, as it is thoroughly
explained in Ref. \cite{koukichtstability}, Eq. (\ref{conz}) can
be written as \beq\frac{\partial\la H_1\ra }{\partial \phi_i}=0
\qquad \Leftrightarrow \qquad \sum_{m=1}^\infty
mA_{k_im}A_{k_{i-1}m}\sin m \phi_i=0\label{ex_gen_2}\eeq where
$A_{j}$ is the $j_{th}$ Fourier coefficient of the single oscillator. From Eq. (\ref{ex_gen_2}), we conclude that
$\phi_i=0$ and $\phi_i = \pi$ always satisfy (\ref{conz}) while, if
special symmetry conditions hold, one could also obtain additional
solutions which would be time non-reversible, following the
terminology of \cite{MacKay1}.

If the action-angle canonical transformation is known, one could
search for these solutions in (\ref{conz}) or its equivalent
(\ref{ex_gen_2}). However, in the generic case where the explicit
form of the action-angle variables is \emph{not} known, a method
to calculate the necessary quantities has been developed in Ref.
\cite{kouk}. According to this method, the system of equations
(\ref{conz}) is equivalent to the following one:
\beq\int_0^T\frac{\pa H_1}{\pa x_i}p_i\ud t=0 \, , \qquad \qquad
i=1\ldots n \, .\label{conxp}\eeq This system can easily be solved numerically,
as will be later shown in a specific example.

Besides the existence of the multibreather-solutions, the phase
difference between the oscillators also determines its linear
stability, as shown in Refs. \cite{arc,koukichtstability}.

The linear stability of a periodic orbit (which in the specific
case is the anticipated multibreather), is defined by the
eigenvalues of the corresponding  Floquet matrix $\lambda_i$ (called also Floquet multipliers), see e.g. \cite{Aubry}. If all these eigenvalues lie on the unit circle of the complex plane then the periodic orbit is linearly stable, while if an eigenvalue lies outside the circle the orbit is unstable. Note that for every eigenvalue we also have its reciprocal and its complex conjugate, because of the Hamiltonian structure of the system (i.e.\ for every $\lambda_i$, there are also $\lambda_i^{-1}$, $\lambda_i^*$ and ${\lambda_i^*}^{-1}$). So, we cannot have just one eigenvalue outside the unit circle but only an even number of them. For
$\e=0$, the above mentioned eigenvalues lie in two complex conjugate bundles at
$e^{\pm i\w_pT_b}$, except the $2n+2$ eigenvalues which correspond
to the $n+1$ central oscillators which lie at unity. For
$|\e|\neq0 \ll 1$, the eigenvalues of the non-central oscillators
move along the unit circle being of the same Krein kind, forming
this way the phonon band, while the ones of the central
oscillators are given by \beq \lambda_{i}=e^{\sigma_iT} \, ,\eeq
where $\sigma_i$ are the corresponding $2n+2$ {\it characteristic
exponents}. As it was proven in Ref. \cite{ams} (and also stated,
within the present formalism, in Ref. \cite{koukicht1}), these
exponents are given in the leading order of approximation by
\beq\sigma_i^2=\e\sigma_{i1}^2+O(\e^{3/2}) \, , \label{sigma}\eeq
where $\sigma_{i1}^2$ coincide with the $n+1$ eigenvalues of the
stability matrix
$$E=-A\cdot B \, , $$
with
$$A=\left(\begin{array}{cc}
\ds\frac{\pa^2\avh}{\pa\thet_i\pa\thet_j}
\end{array}\right),\qquad
B=\left(\begin{array}{cc}
\ds\frac{\pa^2H_0}{\pa J_i\pa J_j}
\end{array}\right).
$$
Therefore, if the various values of $\sigma_{i1}^2$, i.e. the
eigenvalues of $E$, are distinct, and the product $\e\sigma_{i1}^2$ is negative, then the corresponding characteristic exponents $\sigma_i$ are imaginary up to leading order terms and distinct, which results in eigenvalues $\lambda_i$ on the unit circle and the multibreather
is linearly stable. If there are no other solutions than the
standard ones the corresponding linear stability is well defined
by the knowledge of the resonant angles $\phi_i$, the kind of
potential anharmonicity --- i.e. hardening $\left({\pa\w_i}/{\pa
J_i}>0\right)$ or softening $\left({\pa\w_i}/{\pa J_i}<0\right)$
--- and the sign of $\e$. Let us now apply this method in a
specific example, namely the equation of transverse dust grain
motion in a dust crystal.

\section{Transverse particle motion in Debye crystals and Dusty Plasma crystals}

We shall consider the vertical (off-plane, $\sim \hat z$) charged
particle displacement in a crystal (assumed
quasi-one-dimensional, of infinite length: identical grains of
charge $q$ and mass $M$ are situated at $x_i = i\, r_0 , \,$\
where $i= ...,-2, -1, 0, 1, 2, ...$), by taking into account the
intrinsic nonlinearity of the sheath electric (and/or magnetic)
potential. The in-plane (longitudinal, acoustic, $\sim \hat x$ and
shear, $\sim \hat y$) degrees of freedom are assumed suppressed;
this situation is indeed today realized in appropriate experiments
\cite{Liu, Misawa}, where a laser impulse triggers transverse dust
grain oscillations, while a confinement potential ensures the
chain's in-plane stability.

\subsection{Equation of motion}

The vertical grain displacement obeys an equation in the form
\cite{IKPKSTDLWMI, IKPKSTMDLWMI, IKPKSPOPDB, psbook}
\begin{equation}
\frac{d^2 \delta \phi_i}{dt^2} + \nu \frac{d \delta \phi_i}{dt} + \,
\omega_0^2 \, (\,\delta z_{i+1} + \,\delta z_{i-1} - 2 \,\delta
\phi_i) + \omega_g^2 \, \delta \phi_i + \alpha \, (\delta \phi_i)^2 + \beta
\, (\delta \phi_i)^3  = 0 \, , \label{eqmotion}
\end{equation}
where $\delta \phi_i(t) = \phi_i(t) - z_0$ denotes the small
displacement of the $n-$th grain around the (levitated)
equilibrium position $z_0$, in the transverse ($z-$) direction.
The characteristic frequency $\omega_0\,  = \bigl[ - q
\Phi'(r_0)/(M r_0) \bigr]^{1/2}$ results from the dust grain
(electrostatic) interaction potential  $\Phi(r)$, e.g. for a
Debye-H\"uckel potential  \cite{Konopka, Tomme}: \( \Phi_D(r) =
({q}/{r}) \,e^{-{r/\lambda_D}}\), one has: \( \omega_{0, D}^2\,  =
q^2/(M r_0^3) \, (1 + r_0/\lambda_D) \,\exp(-r_0/\lambda_D) \, ,
\) where $\lambda_D$ denotes the effective DP Debye radius
\cite{psbook}. The damping coefficient $\nu$ accounts for
dissipation due to collisions between dust grains and neutral
atoms. The gap frequency $\omega_g$ and the nonlinearity
coefficients $\alpha, \beta$ are defined via the overall vertical
force: $F(z) = F_{e/m} - Mg \approx - M [\omega_g^2 \delta z_n +
\alpha \, (\delta z_n)^2 + \beta \, (\delta z_n)^3 ] \, + {\cal
O}[(\delta z_n)^4]$, which has been expanded around  $z_0$ by
formally taking into account the (anharmonicity of the) local form
of the sheath electric (follow exactly the definitions in Ref.
\cite{IKPKSTDLWMI}, not reproduced here) and/or magnetic
\cite{comment1} field(s), as well as, possibly, grain charge
variation due to charging processes \cite{IKPKSTMDLWMI}. Recall
that the electric/magnetic levitating force(s) $F_{e/m}$
balance(s) gravity at $z_0$. Notice the difference in structure
from the usual nonlinear Klein-Gordon equation used to describe
one-dimensional oscillator chains --- cf. e.g. Eq. (1) in Ref.
\cite{Kivshar}: TDLWs (\textit{`phonons'}) in this chain are
stable only in the presence of the field force $F_{e/m}$.

For convenience, the time and vertical displacement variables may
be scaled over appropriate quantities, i.e. the characteristic
(single grain) oscillation period $\omega_g^{-1}$ and the lattice
constant $r_0$, respectively, viz. $t = \omega_g^{-1} \tau$ and
$\delta z_n = r_0 q_n$; Eq. (\ref{eqmotion}) is thus expressed as:
\begin{equation}
\frac{d^2 q_n}{d \tau^2} + \, \epsilon (\,q_{n+1} + \, q_{n-1} - 2
\,q_n) + \, q_n + \alpha' \, q_n^2 + \beta' \, q_n^3 = 0 \, ,
\label{eqmotion1}
\end{equation}
where the (dimensionless) damping term, now expressed as
$({\nu}/{\omega_g}) {d q_n}/{d\tau} \equiv \nu' \dot{q}_n$, will
be henceforth omitted in the left-hand side (assuming $\nu' \ll
1$). The coupling parameter is now $\epsilon =
{\omega_0^2}/{\omega_g^2}$, and the nonlinearity coefficients are
now: $\alpha' = \alpha r_0/\omega_g^2$ and $\beta' = \beta
r_0^2/\omega_g^2$. All quantities in Eq. (\ref{eqmotion1}) are
dimensionless \footnote{Note, for rigor, that the coupling
strength $\epsilon$ in (\ref{eqmotion1}) is formally the opposite
of $\epsilon$ in (\ref{KG}), hence the inverse-dispersive
character discussed in the text. We shall clarify this subtlety in
notation further below.}.

\subsection{Linear transverse dust-lattice waves}

Retaining only the linear contribution and considering
oscillations of the type, $\delta z_n \sim \,\exp[i \,(k n r_0 -
\omega t)] + c.c.$ (complex conjuguate) in Eq. (\ref{eqmotion}),
one obtains the well known transverse dust lattice (TDL) wave
optical-mode-like dispersion relation
\begin{equation}
\omega^2\,  = \omega_g^2\, - 4 \omega_0^2\, \sin^2 \biggl( \frac{k
r_0}{2}
\biggr)
\, , \label{dispersion-discrete}
\end{equation}
or
\begin{equation}
\tilde \omega^2 = 1\, - 4 \epsilon \, \sin^2 ({\tilde k}/{2}) \, ,
\label{dispersion-discrete1}
\end{equation}

\begin{figure}[htbp]
\begin{centering}
\includegraphics[width=7cm]{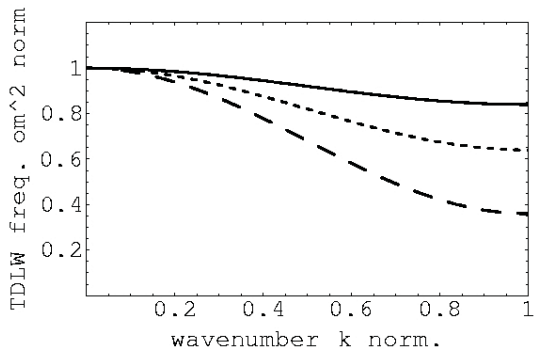}
    \caption{The dispersion relation of the TDL excitations:
frequency $\omega$ (normalized over $\omega_g$) versus wavenumber
$k$. The value of $\omega_0/\omega_g$ ($\sim$ coupling strength)
increase from top to bottom. Note that upper (less steep,
continuous) curve is more likely to occur in a real
(weakly-coupled) DP crystal.}
    \label{disp_diag}
    \end{centering}
\end{figure}

See that the wave frequency $\omega \equiv \tilde \omega \omega_g$
\emph{decreases} with increasing wavenumber $k = 2 \pi/\lambda
\equiv \tilde k/r_0$ (or decreasing wavelength $\lambda$), see Fig. \ref{disp_diag}, 
implying that transverse vibrations propagate as a \emph{backward
wave}: the group velocity $v_g = \omega'(k)$ and the phase
velocity $\omega_{ph}=\omega/k$ have opposite directions (this
behaviour has been observed in recent experiments). The
modulational stability profile of these linear waves (depending on
the plasma parameters) was investigated in Refs.
\cite{IKPKSTDLWMI, IKPKSTMDLWMI}. Notice the natural \emph{gap
frequency} $\omega(k=0)\, = \omega_g = \omega_{max}$,
corresponding to an overall motion of the chain's center of mass,
as well as the \emph{cutoff frequency} $\omega_{min}\, =
(\omega_g^2\, - 4 \omega_0^2)^{1/2} \equiv \omega_g\, (1 - 4
\epsilon^2)^{1/2}$ (obtained at the end of the first Brillouin
zone $k = \pi/r_0$ \cite{Kittel}) which is \emph{absent in the continuum limit},
viz. $\omega^2\, \approx \omega_g^2\, - \omega_0^2 \, k^2 \,
r_0^2$ (for $k \ll r_0^{-1}$); obviously, the study of wave
propagation in this ($k \lesssim \pi/r_0$) region invalidates the
continuum treatment employed so far in literature. The essential
feature of discrete dynamics, to be retained here, is the (narrow)
bounded TDLW (\textit{`phonon'}) frequency band, limited in the
interval $\omega \in [(\omega_g^2\, - 4 \omega_0^2)^{1/2},
\omega_g]$; note that one thus naturally obtains  the stability
constraint: $\omega_0^2/\omega_g^2 = \epsilon < 1/4$ (so that
$\omega \in \mathbb{R} \quad \forall k \in [0, \pi/r_0]$).

We needn't go into further details concerning the linear regime,
since it is covered rather extensively in the literature
\cite{psbook}. We shall, instead, see what happens if the {\em
nonlinear} terms are retained, in this discrete description.

\subsection{Multibreathers in Debye crystals and Dusty Plasma crystals}

Eq.  (\ref{eqmotion1}) can be generated by a Hamiltonian of the
form (\ref{hamfull}) by considering a quartic on-site (sheath)
polynomial potential of the form \beq V(x_i)=\frac{1}{2}x_i^2 +
\frac{\alpha'}{3}x_i^3 + \frac{\beta'}{4} x_i^4 \, , \eeq and considering
\emph{negative} values of $\e$ (in account of inverse dispersion).
Note that, henceforth $x$ stands for $q$ of (\ref{eqmotion1}),
i.e.\, describes the normalised vertical displacement from the
equilibrium.

The values of  the anharmonicity parameters $\alpha'$ and $\beta'$
may be deduced from dusty plasma experiments on nonlinear vertical
dust lattice oscillations \cite{Ivlev2000,Zafiu,Misawa,Liu}. For
instance, the Kiel (Germany) experiment by Zafiu \textit{et al.}
\cite{Zafiu} -- using a laser to trigger nonlinear vertical dust
grain oscillations -- has provided the values: $\alpha/\omega_g^2
= + 0.02; \, +0.016; \, -0.27 \, ({\rm{mm^{-1}}})$ and
$\beta/\omega_g^2 = -0.16; \, -0.17; \, -0.03 \, ({\rm{mm^{-2}}})$
(successively, by gradually increasing the diameter of the dust
grains; see Table I in Ref. \cite{Zafiu}). In our notation, this
implies: $\alpha' \simeq + 0.02; \, +0.016; \, -0.27$,  and
$\beta' \simeq -0.16; \, -0.17; \, -0.03$ (for a lattice spacing
of the order of $r_0 \simeq 1 \, {\rm{mm}}$). Note that damping
was very low ($\nu' \simeq 0.02$), thus a posteriori justifying
our neglecting it.

The experiment on anharmonic \emph{single} grain oscillations by
Ivlev \textit{et al.} \cite{Ivlev2000}, carried out in Garching
(Germany), provides curve fitting data for $\Phi(z)$, i.e.
$\alpha/\omega_g^2 = - 0.5 \, {\rm{mm^{-1}}}$ and
$\beta/\omega_g^2 = 0.07 \, {\rm{mm^{-2}}}$. One thus deduces
$\alpha' \approx - 0.5 $ and $\beta' \approx 0.07$ (for a lattice spacing,  say typically, of the order of $r_0 \approx 1 {\rm{mm}}$). Note that the damping coefficient $\nu$ was
 as low as $\nu/2\pi \simeq 0.067 \, {\rm{sec}}^{-1}$, so that (with $\omega_g/2\pi
 \simeq 17 \, {\rm{sec}}^{-1}$) one has: $\nu' = \nu/\omega_g \simeq
 0.004$ (the pressure in that
experiment was  kept as low as 0.5 Pa; see the original paper for
technical details on the experimental device).  The sets of anharmonicity values are shown together in Table
\ref{vartable}, for reference.

\begin{table}[!htb]
\begin{tabular}{| c | c | c | c | c |}
\hline
&\multicolumn{3}{|c|}{\begin{tabular}{c}Kiel\\[-4pt]Experiment\end{tabular}}&\begin{tabular}{c}Garching\\[-4pt]Experiment\end{tabular}\\
\hline
&\rm{I}&\rm{II}&\rm{III}&\rm{IV}\\
\hline
$ \alpha' $& 0.02 & 0.016 & -0.27 & -0.5\\
\hline
$\beta'$&-0.16&-0.17&-0.03&0.07\\
\hline
\end{tabular}
\caption{Experimental data: Sets I, II, and III of sheath
potential anharmonicity values, as obtained from the data in Ref.
\cite{Zafiu}. Column IV provides the data in Ref. \cite{Ivlev2000}.} \label{vartable}
\end{table}

The results of the  experiment on linear TDLWs by Misawa et al.
\cite{Misawa} allows for a rough estimation of the coupling
strength (still for low pressure; see details in Ref.
\cite{Misawa}): $\omega_g \simeq 155 \, {\rm{sec}}^{-1}$ and
$\omega_0 \simeq 19.5 \, {\rm{sec}}^{-1}$ (derived from Fig. 3a
therein), which give $\epsilon \simeq 0.016$. Note that the
effective damping term was kept as low as $\nu \simeq 0.239 \,
{\rm{sec}}^{-1}$, i.e. $\nu' = \nu/\omega_g \simeq
 0.00154$.

\begin{figure}[htbp]
\begin{centering}
\begin{tabular}{ccc}
\includegraphics[width=5.5cm]{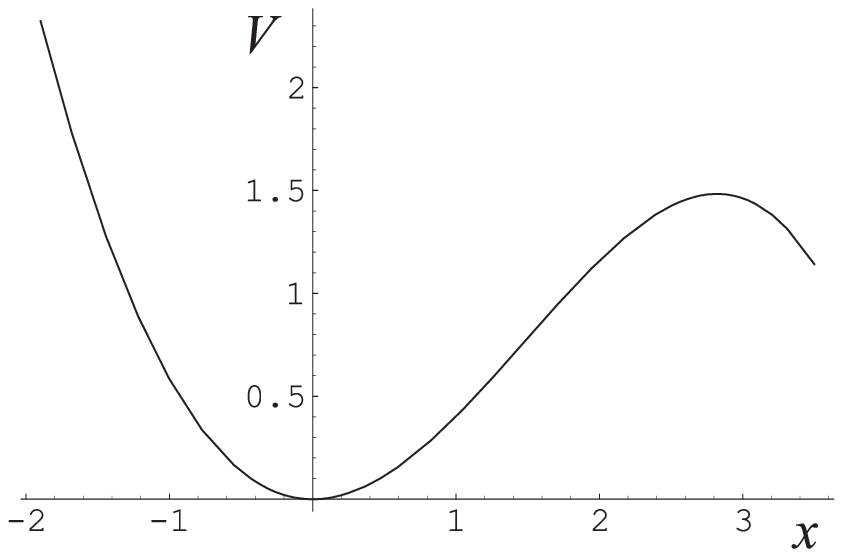}&\includegraphics[width=6cm]{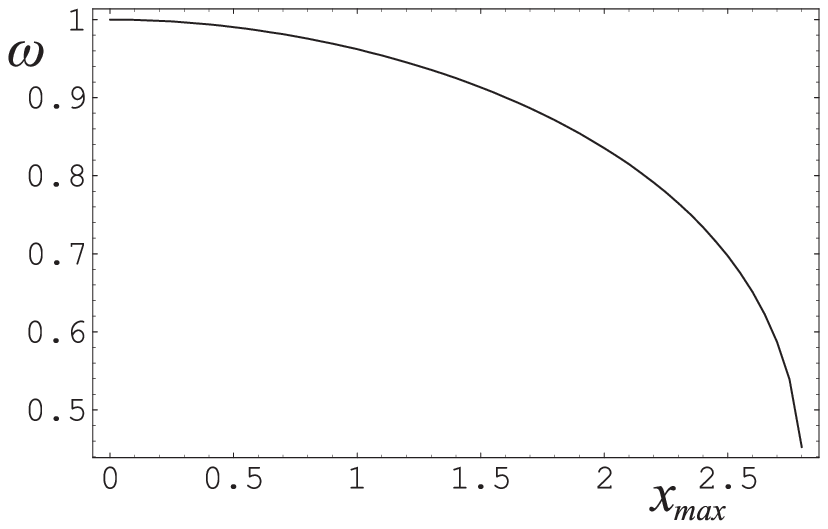}&\includegraphics[width=6cm]{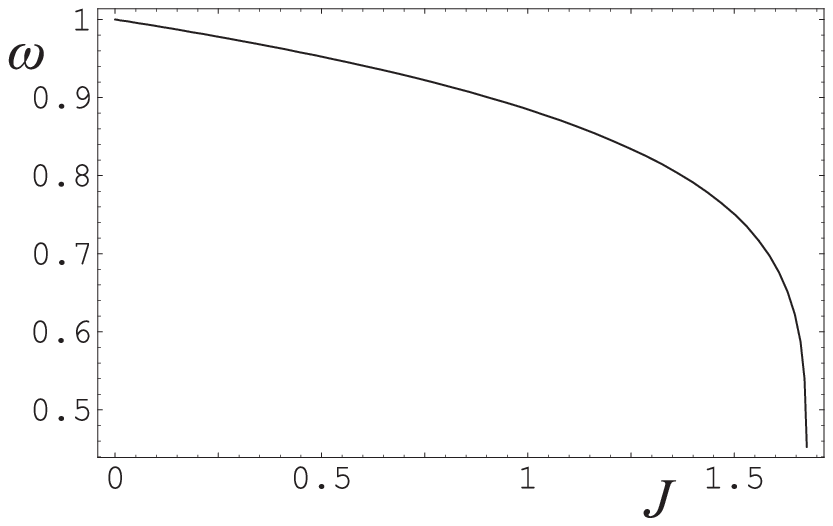}\\[-10pt]
(a)&(b)&(c)
\end{tabular}
    \caption{(a) The on-site potential is depicted against the vertical displacement $x$. (b) The frequency $\w$ is depicted against the amplitude of the oscillation $x_{max}$. (c) The frequency $\w$ is depicted against the action variable $J$. All the figures have been obtained by using the set of values III in Table \ref{vartable}.}
    \label{fig1}
    \end{centering}
\end{figure}

We will first examine the system provided by Zafiu \textit{et al.}
in \cite{Zafiu}. Since the first three sets of values presented in Table \ref{vartable} provide similar results we shall
consider only the results in the third column, i.e.\ $\alpha'=-0.27$ and $\beta'=-0.03$. The on-site potential $V(x)$ as a function of the vertical displacement $x$ is depicted in Fig.\ref{fig1}(a). The frequency $\w$ dependence on the maximum displacement of the free oscillator $x_{max}$ is shown in Fig. \ref{fig1}(b), while its dependence on the action $J$ is shown in Fig. \ref{fig1}(c). Note that, since there is a one to one correspondance between $x_{max}$ and $J$, figs \ref{fig1}(b) and \ref{fig1}(b) are equivalent. Due to the form of the potential $V(x)$, the region of motion
is restricted to the region from $x\simeq-1.55$ to $x \simeq 2.82$.
Let us remark, in passing, that this potential form essentially
accounts for crystal dissociation (melting), which occurs when
vibrating particles overcome the finite potential barrier. We note that $\omega(J)$ is a decreasing function of
$J$, so the anharmonicity condition $\partial\omega/\partial
J\neq0$ is satisfied everywhere in the allowed region of motion.
The computations of $\w(x)$ and $\w(J)$ has been made numerically since the
explicit action-angle transformation is not known; for a more
detailed description, see Ref. \cite{kouk}. In Fig.\ref{fig1}, $\w(J)$ together with the potential function of the specific choice of coefficients are shown. 
We consider the anticontinuous limit of only two oscillators moving
in resonant periodic orbits with frequencies $\w_1$ and $\w_2$ and
the rest in the equilibrium. We examine the 1:1 and 2:3 cases i.e.
the cases with $\w=\w_1=\w_2$ and
$\w=\frac{\w_1}{2}=\frac{\w_2}{3}$. We are thus simply left with the task of choosing appropriate values of the amplitude $x_{max}$, corresponding to the two values of the frequency $\w_1$ and $\w_2$, which should fulfil the resonance condition and, at the same time, the frequency $\w$ should not be {\it close to} the harmonic vibration frequency $\w_p$ (here 1, i.e.\ the linear frequency in the absence of coupling). This is easily accomplished upon simple inspection of Fig. \ref{fig1}(b) or \ref{fig1}(c). Note that we have to avoid a neighborhood of values around $\w_p$, since for $\e\neq0$ a phonon band will be formed instead of the single frequency $\w_p$ and possible resonance of the breather frequency with this band will end up in energy loss via the excitation of linear waves, so that localised discrete modes will be destabilized. By fixing the orbits of the two oscillators we determine at the
same time the energy of the orbits at the anticontinuous limit as
well as the period $T$ of the produced multibreather. We now have
to check  which phase difference the central oscillators must have
in order to form a multibreather for $\e\neq0$.

\begin{figure}[htbp]
\begin{centering}
\begin{tabular}{ccc}
\includegraphics[width=7cm]{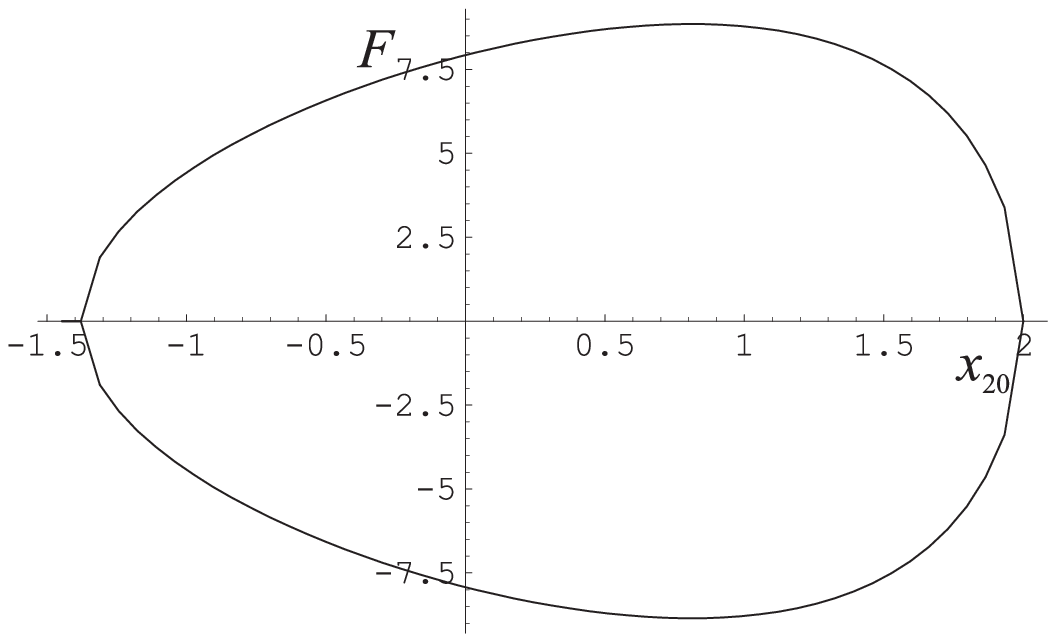}&\hspace{1cm}&\includegraphics[width=7cm]{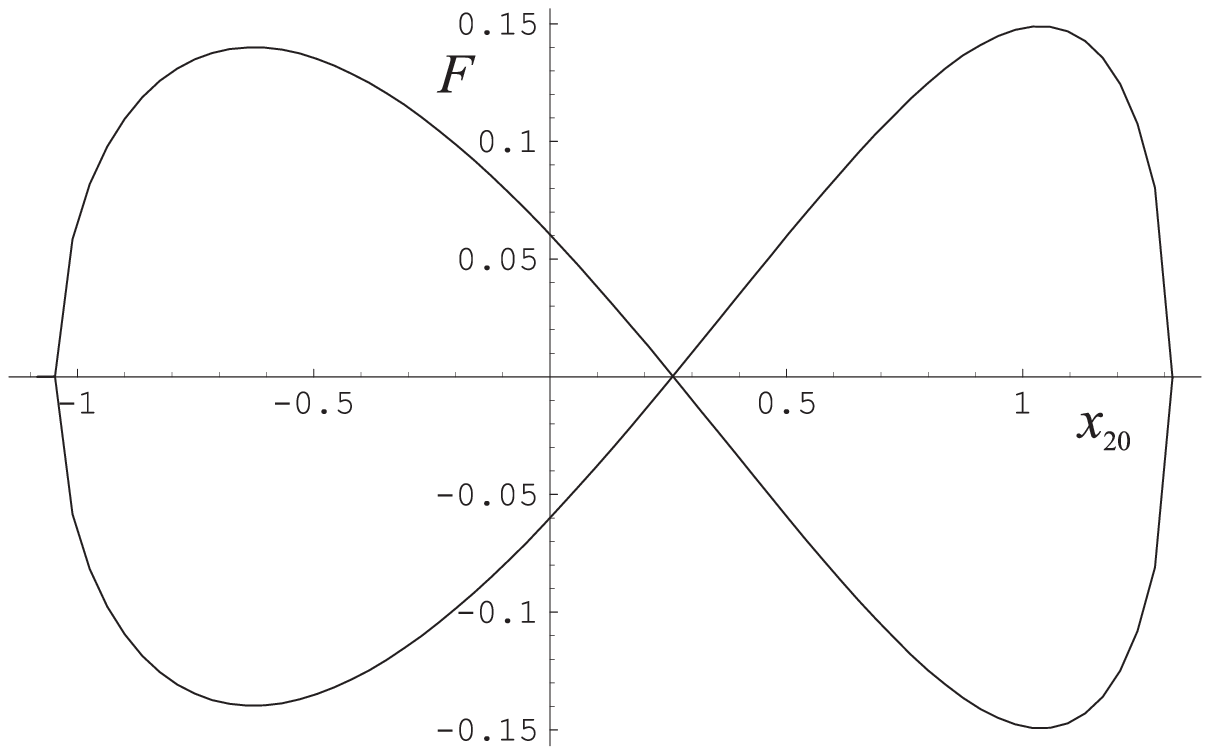}\\
(a)&\hspace{1cm}&(b)\\
\end{tabular}
    \caption{$F(x_{20})$ for the 1:1 and 2:3 resonances.}
    \label{fig2}
    \end{centering}
\end{figure}

We now have to check condition (\ref{conz}). Since we consider two oscillators in the anticontinuous limit ther is only one angle $\phi$ and, as already shown
above, the $\phi=0,\pi$ solutions always exist. Since the
action-angle transformation is not analytically known, in order to
check for extra solutions, one has to solve the equivalent
equation (\ref{conxp}), i.e., for $i=2$, 
\beq\int_0^T\frac{\pa H_1(x_1,x_2)}{\pa
x_2}p_2\ud t=0 \, .\label{conxp2}\eeq Note that all of the above calculations
are made in the anticontinuous limit ($\epsilon = 0$). Since the periodic orbits are
given, the above equation defines a relationship between the
initial conditions of the two orbits, $x_{10}$, $p_{10}$,
$x_{20}$, $p_{20}$. We fix $x_{10}=0$ and choose the value of
$p_{10}>0$ which is specified by the periodic orbit. So, the only
free variable is $x_{20}$, since $p_{20}$ can be accordingly
calculated from the equation of energy. Eq.\ref{conxp2} now leads to \cite{kouk}: \beq F(x_{20})=\int_0^Tx_1p_2\ud t=0 \, .\eeq This
equation is two branched, i.e. yields one branch for each choice
of sign for the momentum $p_{20}$. In Fig. \ref{fig2}, these two
branches are presented together in the same diagram for the 1:1 as
well as for the 2:3 resonance.

For the 1:1 resonance we see that $F(x_{20})$ has two roots, which
correspond to the standard breather solutions $\phi=0,\pi$. As for
the stability of these solutions, following the arguments in Refs.
\cite{arc,koukichtstability}, the solution with $z=0$ will be the
linearly stable one among these two and, since there are no other
solutions besides the ones already mentioned, this will be the only
linearly stable solution. In particular, in Ref.
\cite{koukichtstability} it is shown that $\sigma_{11}$ in
(\ref{sigma}) is given by \beq
\sigma_{11}^2=-\frac{\partial\w}{\partial J}\sum_{m=1}^\infty
m^2A_{m}^2\cos m \phi \, ,  \label{sigma11}\eeq which for $\phi=0$, $\frac{\partial\w}{\partial J}<0$ and $\e<0$ provides imaginary characteristic exponents $\pm\sigma_{1}$ up to $O(\sqrt{\e})$ and consequently linearly stable multibreather, which confirms what has been claimed above. Note that Eq. (\ref{sigma11}) is essentially equivalent (for $\phi=0$ and 1:1 resonance) to the simple 
stability criterion 
\begin{equation}
\e\frac{\pa\w}{\pa J}>0 \, . \end{equation}

We have computed this solution for only two central oscillators,
but the procedure would be exactly analogous for any number $n$ of central oscillators
since, as shown in \cite{kouk}, the system consists of
independent equations. In that case, the only linearly stable
solution would be $\phi_i=0$, for $i=1, \ldots , n$.

\begin{figure}[htbp]
\begin{centering}
\begin{tabular}{ccc}
\includegraphics[width=4cm]{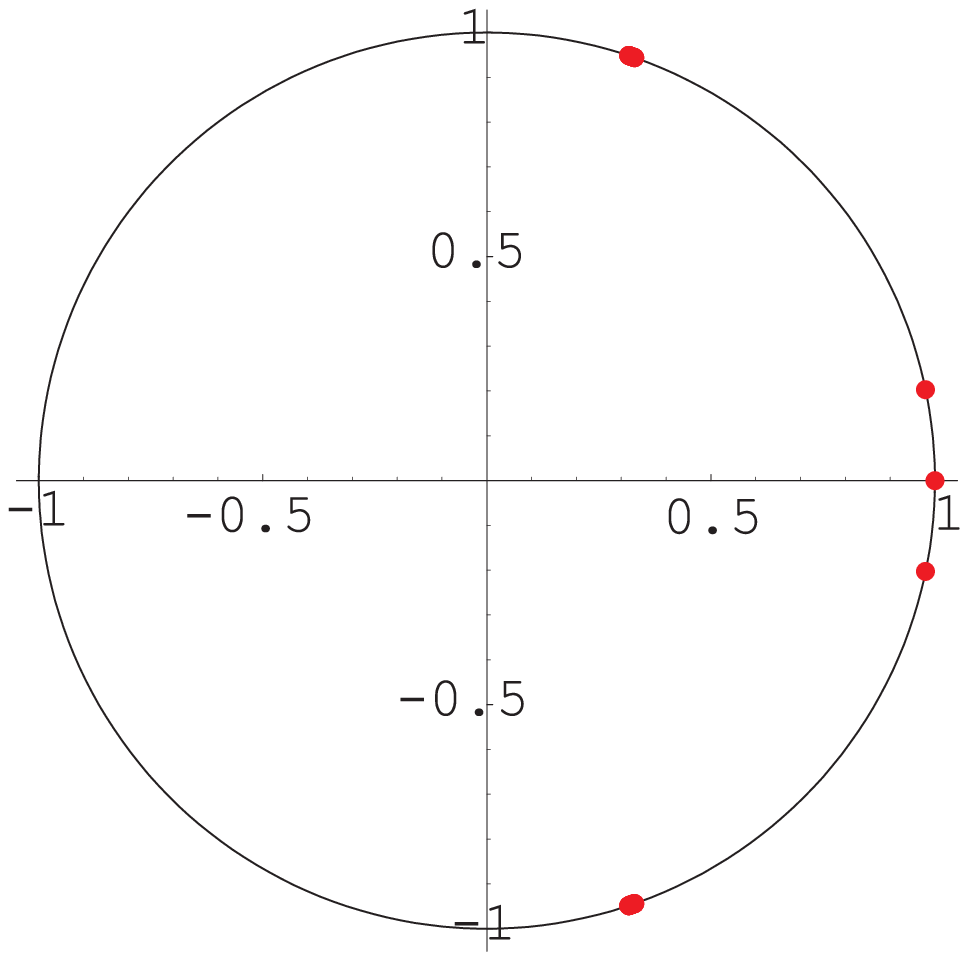}&\hspace{0.5cm}\includegraphics[width=4cm]{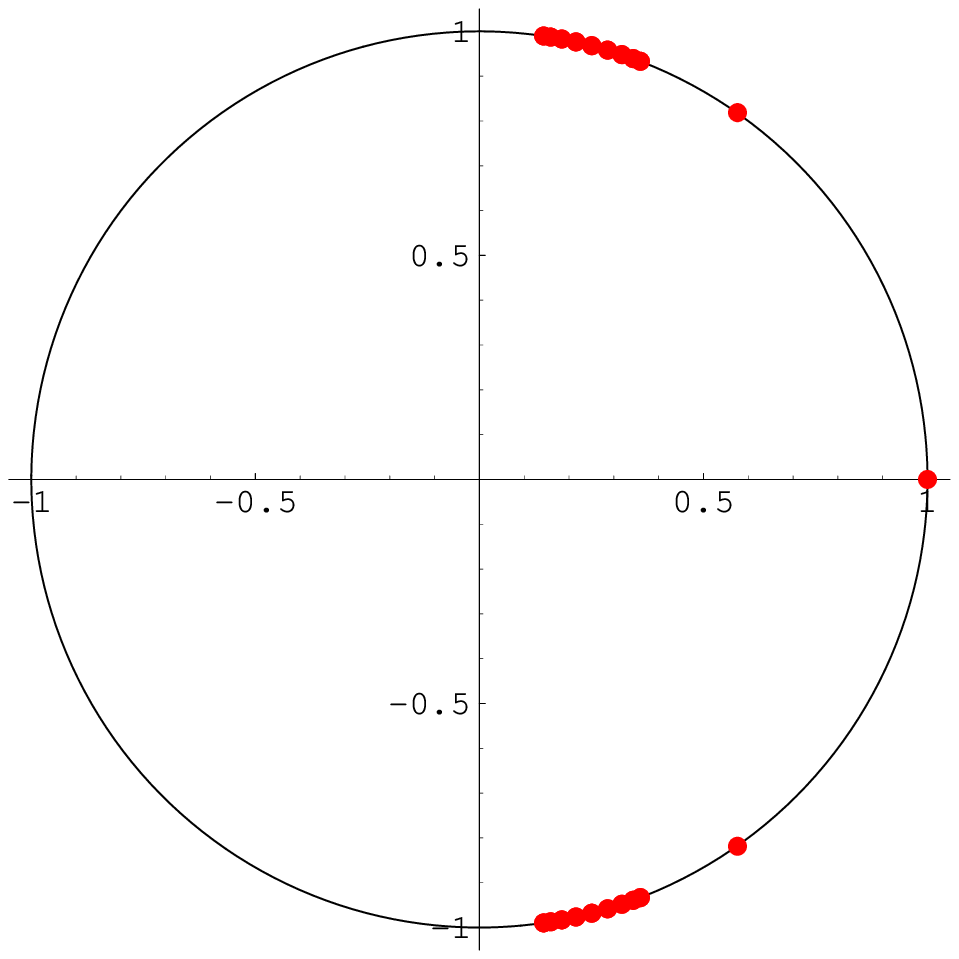}&\hspace{0.5cm}\includegraphics[width=4cm]{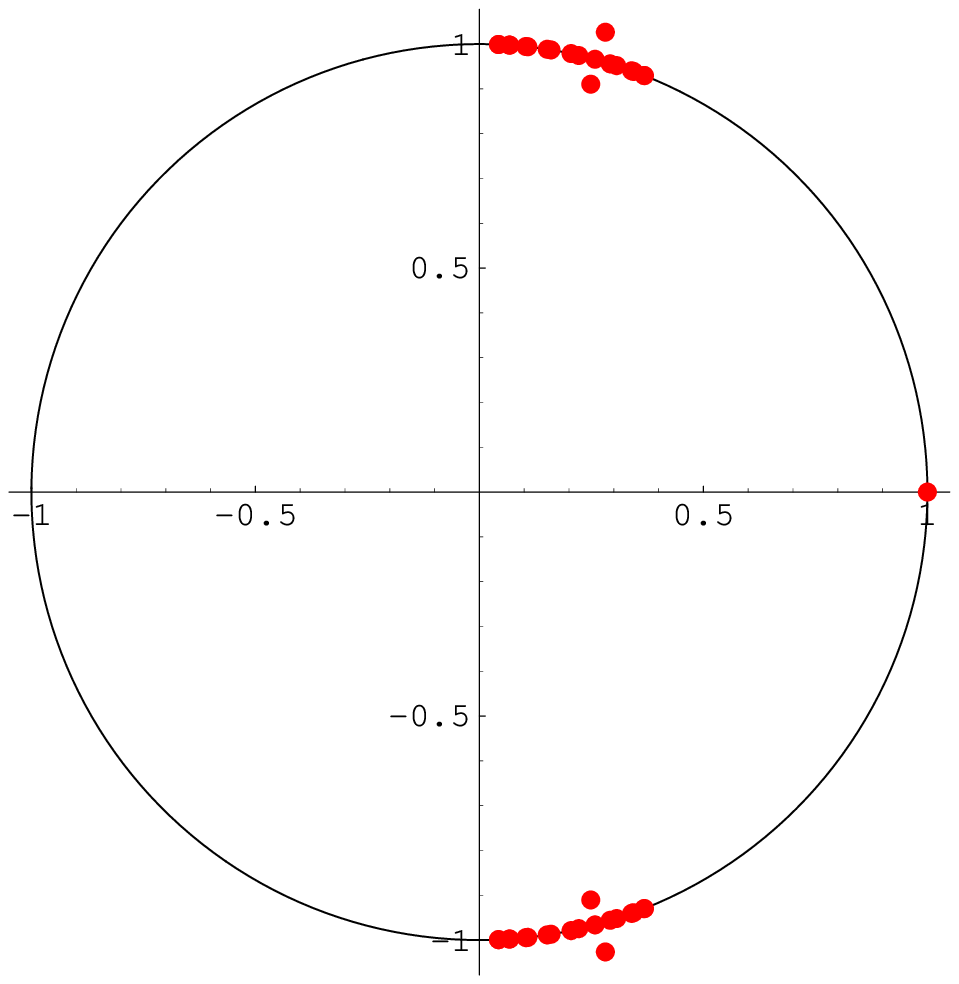}\\
$\e=-0.001$&\hspace{0.5cm}$\e=-0.016$&\hspace{0.5cm}$\e=-0.022$\\
\includegraphics[width=4cm]{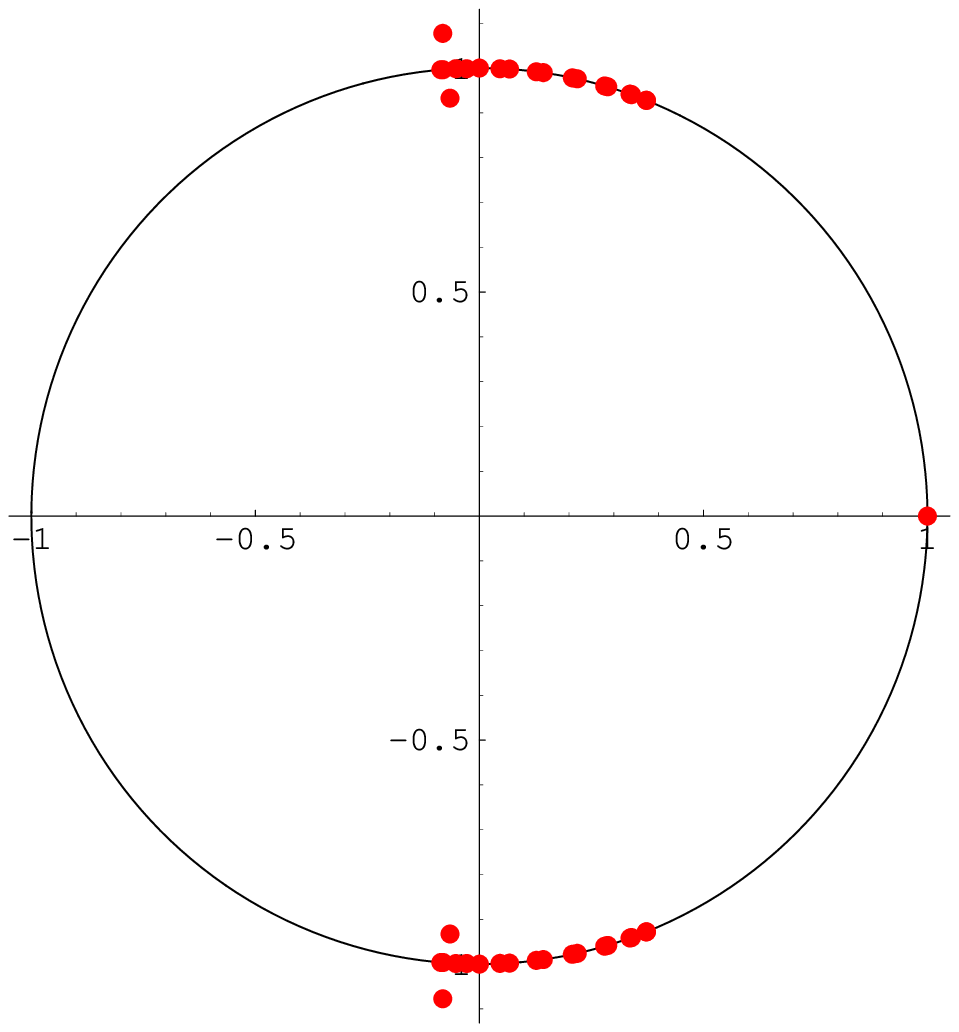}&\hspace{0.5cm}\includegraphics[width=4cm]{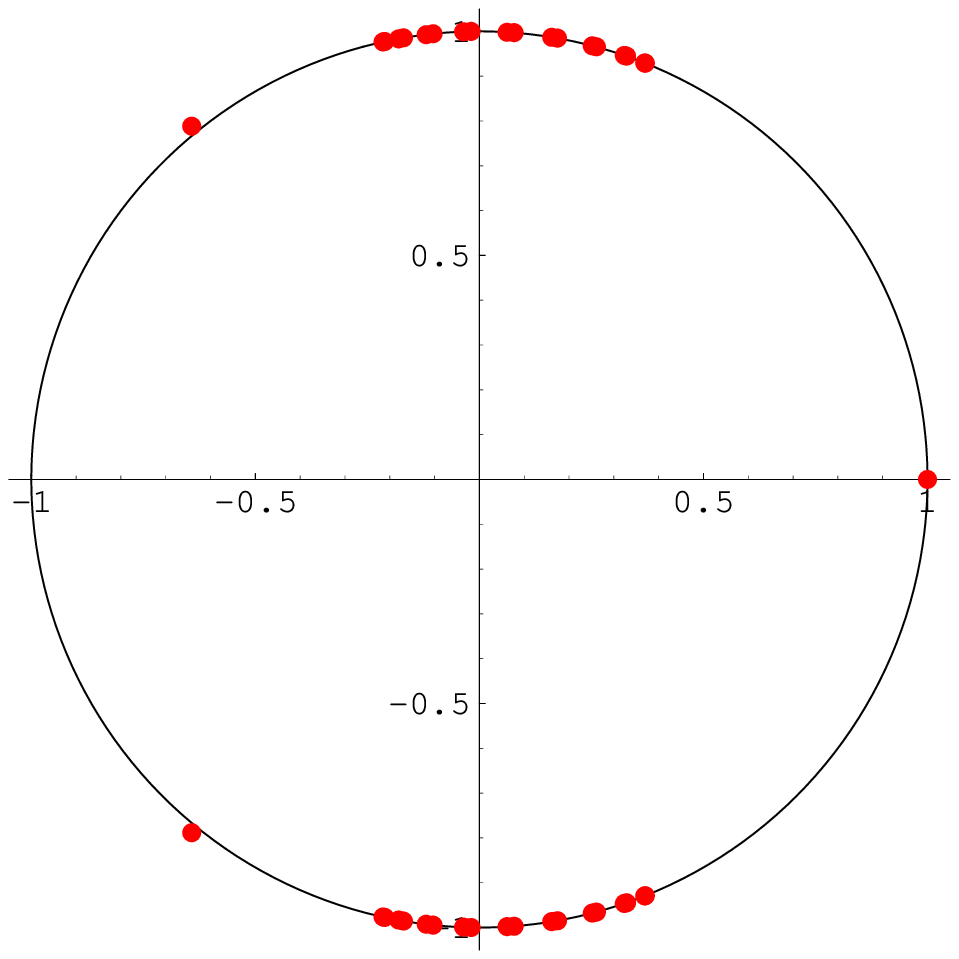}&\hspace{0.5cm}\includegraphics[width=4cm]{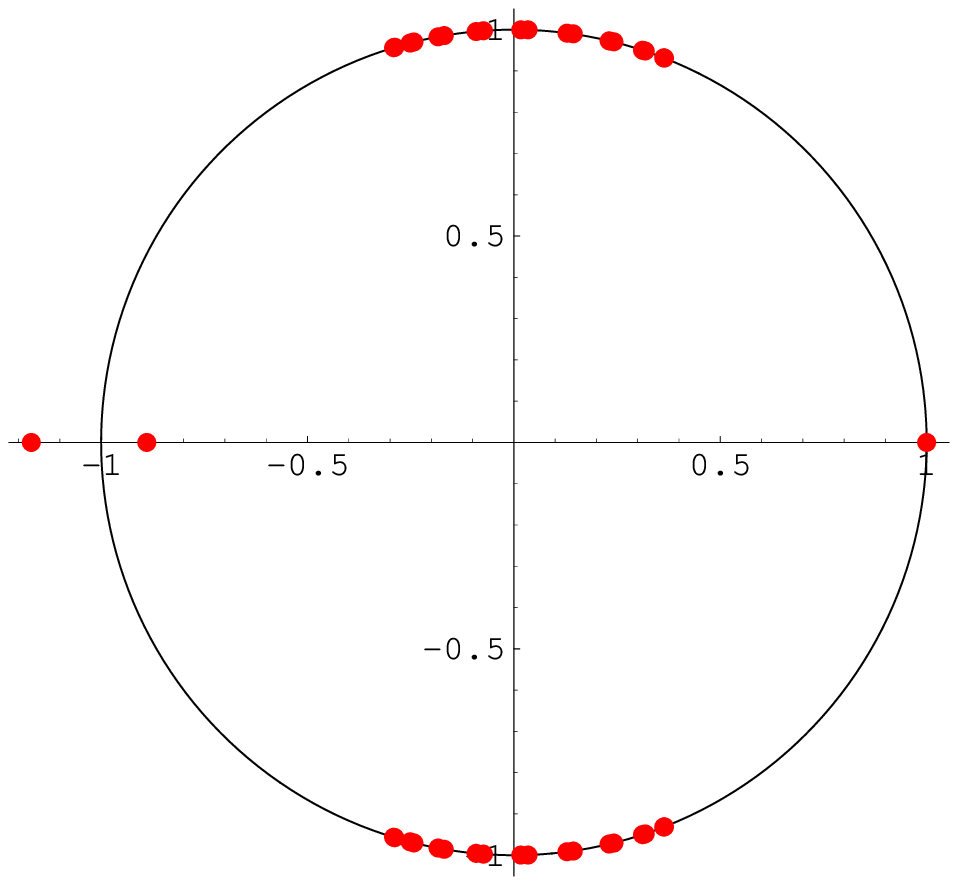}\\
$\e=-0.030$&\hspace{0.5cm}$\e=-0.037$&\hspace{0.5cm}$\e=-0.041$\\\end{tabular}
    \caption{The destabilisation scenario represented by the eigenvalues of the Floquet matrix for growing values of $|\e|$.}
    \label{fig3}
    \end{centering}
\end{figure}

The above mentioned solution is proven to be linearly stable for
values of the coupling parameter $\e$ close enough to zero. However, as the absolute value of $\e$
increases, the central oscillators eigenvalues move along the unit circle, while the non-central oscillators eigenvalues will also move along the oposite direction, forming this way the phonon band.  As the value of $|\e|$ is increasing, at some point the central oscillator eigenvalue will penetrate the phonon band and, since they are of
opposite Krein sign, they can leave the unit circle forming a
complex quadruple (see e.g.\ \cite{Aubry}); the multibreather thus becomes unstable, since there will be four eigenvalues outside the unit circle. By
further increasing $|\e|$ the eigenvalues will return to the unit
circle re-stabilizing the multibreather, which remains stable
until the eigenvalues collide at $-1$ and leave the unit circle
along the real axis causing a period doubling bifurcation and
consequently, the final destruction of the breather. This
destabilisation scenario is shown in Fig.\ref{fig3}.

\begin{figure}[htbp]
\begin{centering}
\includegraphics[width=11cm]{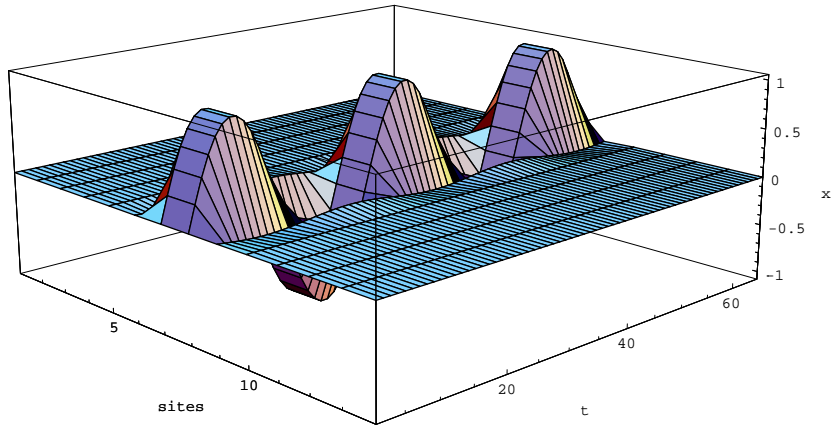}
    \caption{Time evolution of a 1:1 2-breather for $\e=-0.016$.}
    \label{fig4}
    \end{centering}
\end{figure}

Anticipating an experimental confirmation of our results, the
experimentally measured value of the coupling constant $\e$ should
be inside the stability regions. By choosing the amplitude of the
oscillation at the anticontinuous limit to be $x_{max}\simeq1.51$,
a stable multibreather is achieved for $\e=0.016$, as it can be
seen in Fig.\ref{fig3}; this fact confirms that this kind of
motion can be supported by the specific model. The time-evolution of the resulting
2-breather is depicted in Fig.\ref{fig4}.

\begin{figure}[htbp]
\begin{centering}
\includegraphics[width=11cm]{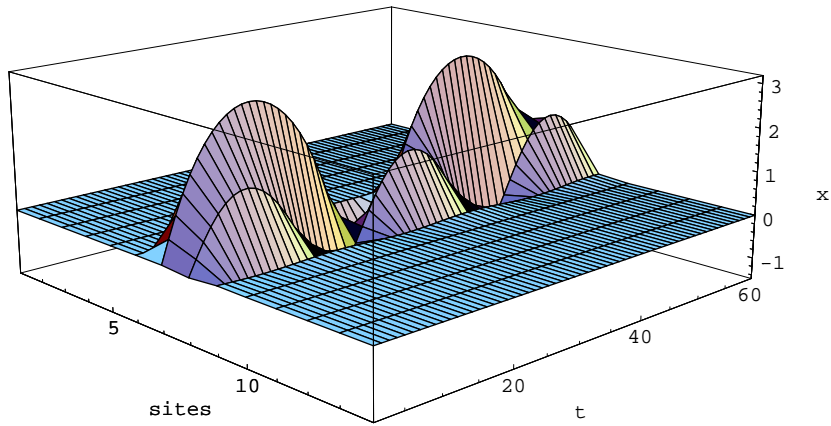}
    \caption{Time evolution of a 2:3 2-breather for $\e=-0.003$.}
    \label{fig5}
    \end{centering}
 \end{figure}

Let us examine the 2:3 resonance now. Note that, in Fig.
\ref{fig2} the curve $F(x_{20})$ crosses the $x$-axis four times.
These four roots correspond to only two values of $\phi$ only,
namely the trivial ones $\phi=0$ and $\pi$.  The resulting
multibreathers, which have the form of Fig.\ref{fig5}, are very quickly (for small values of $\e$)
destabilized so they are physically meaningless and will not be
examined any further.

\begin{figure}
\begin{centering}
\begin{tabular}{ccccc}
\includegraphics[width=5.5cm]{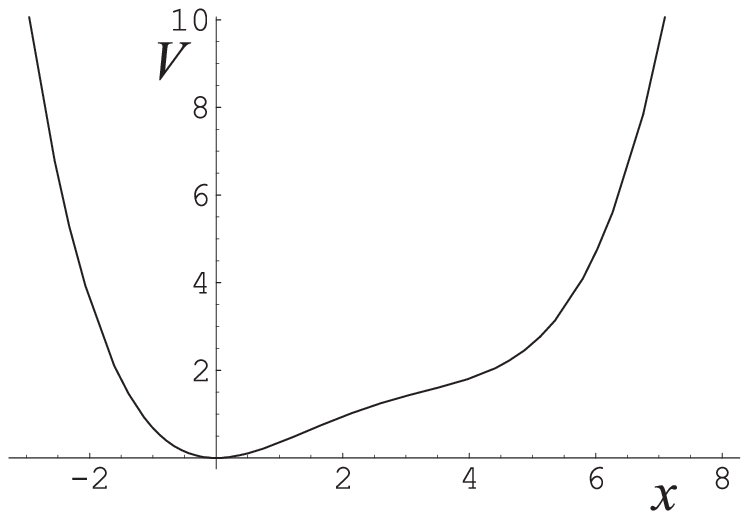}&\hspace{1cm}&\includegraphics[width=6cm]{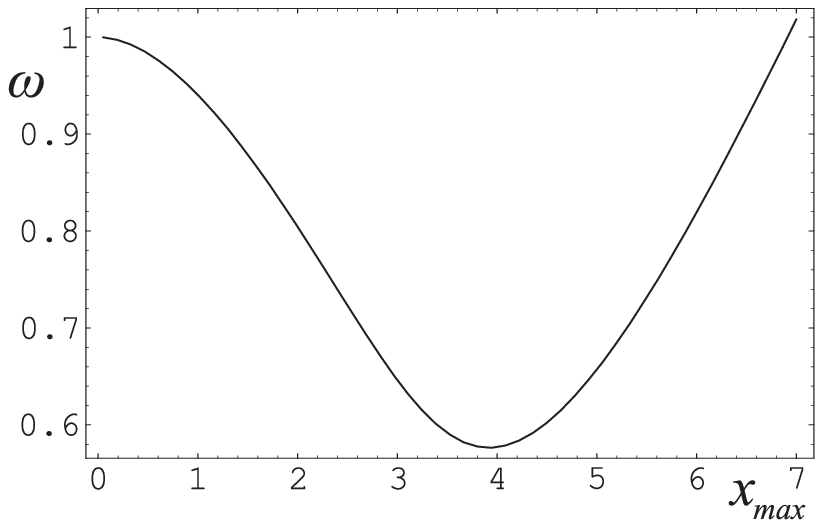}&\hspace{1cm}&\includegraphics[width=6cm]{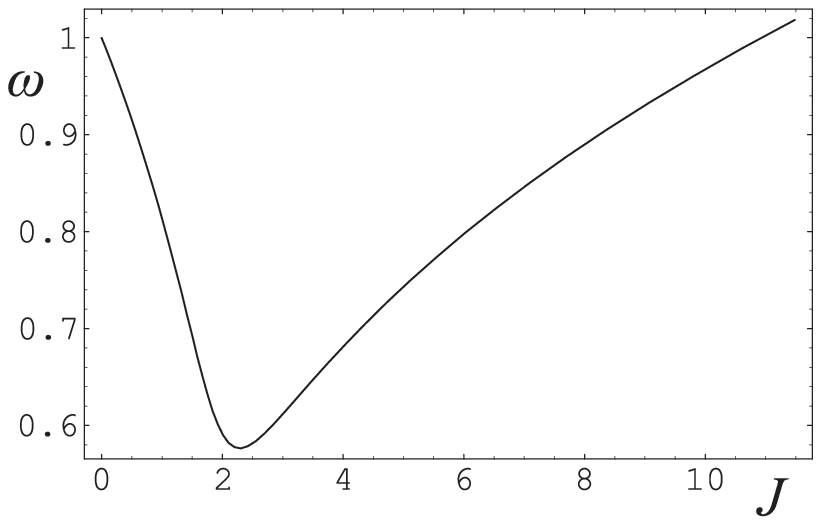}\\[-10pt]
(a)&\hspace{1cm}&(b)&\hspace{1cm}&(c)
\end{tabular}
    \caption{(a) The on-site potential is depicted against the vertical displacement $x$. (b) The frequency $\w$ is depicted against the amplitude of the oscillation $x_{max}$. (c) The frequency $\w$ is depicted against the action variable $J$. All the figures have been obtained using the set of values IV in Table \ref{vartable}.}
    \label{fig6}
    \end{centering}
\end{figure}

\begin{figure}[htbp]
\begin{centering}
\begin{tabular}{ccc}
\includegraphics[width=6cm]{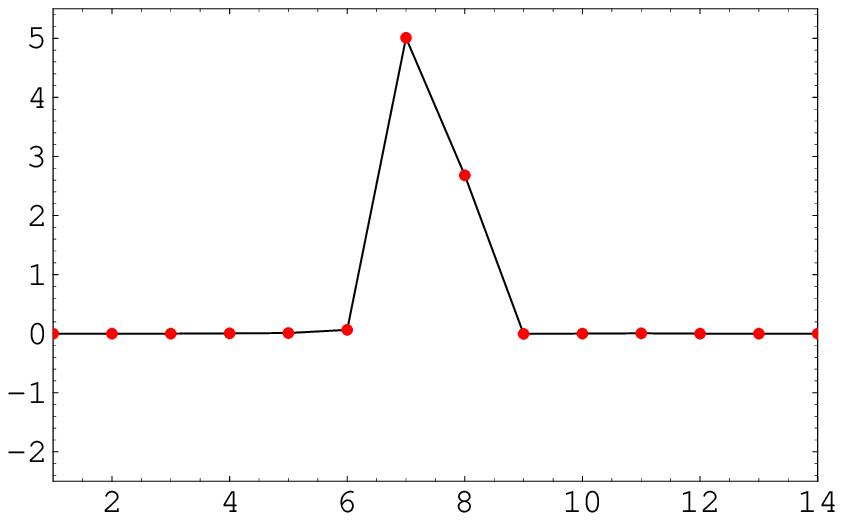}&\hspace{0.5cm}&
\includegraphics[width=6cm]{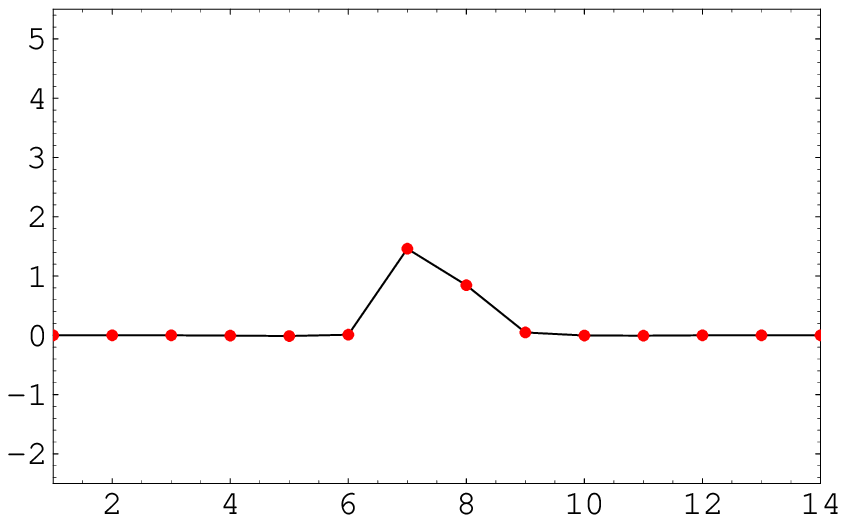}\\
$t=t_1$&&$t=t_2$\\[8pt]
\includegraphics[width=6cm]{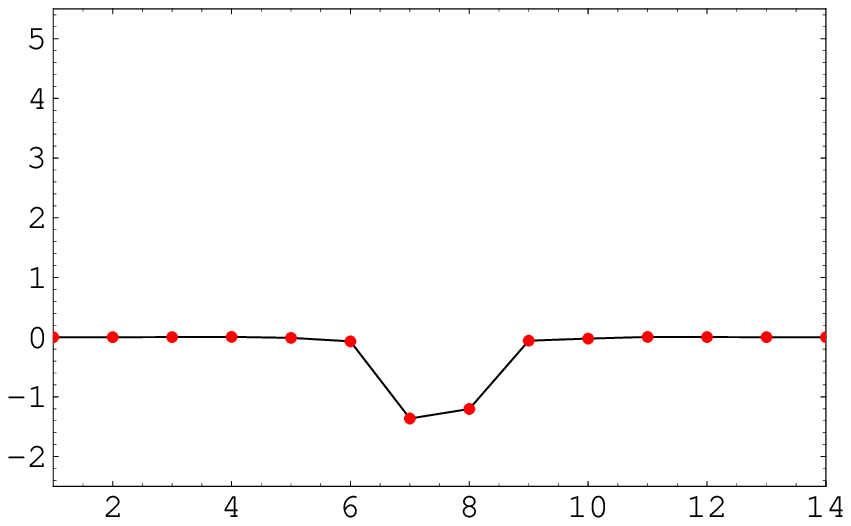}&\hspace{0.5cm}&
\includegraphics[width=6cm]{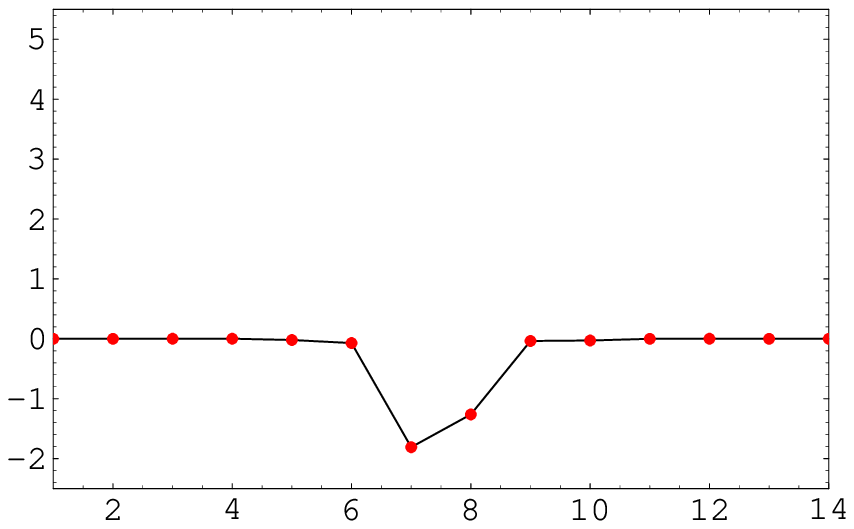}\\
$t=t_3$&&$t=t_4$
\end{tabular}
    \caption{Snapshots of an nonsymmetric multibreather for 4 successive time moments.}
    \label{fig7}
    \end{centering}
\end{figure}

Consider now the nonlinearity values given by Ivlev \textit{et
al.} in \cite{Ivlev2000}. The corresponding potential and
frequency functions are shown in Fig.\ref{fig6}. In this case, the
frequency $\w(J)$ is not everywhere a monotonous function of $J$
but presents a minimum at $J\simeq 2.3$. If we constrain the
allowed region of motion upto the minimum of $\w$, then we obtain
similar results to the above mentioned. So, we can obtain stable
$1:1$ symmetric multibreathers. However, if we extend the allowed
region of motion, we can consider a $1:1$ anticontinuous limit with
the oscillators moving with different amplitudes but the same
frequency. This limit is continued for $\e\neq0$ to provide a
non-symmetric $1:1$ 2-breather as it is shown in Fig.\ref{fig7}.

\section{Conclusions}

This study was devoted to an investigation, from first principles,
of the existence of discrete multi-site lattice excitations
(multibreathers) in a Debye (e.g., dusty plasma) crystal. Relying
on the analytical and numerical methodology established in Refs. \cite{koukicht1,koukichtstability, kouk},
we have shown that Debye crystals may, indeed, support multi-site highly localized oscillatory motion (multibreathers). In particular, employing nonlinearity
parameters provided by plasma discharge experiments \cite{Zafiu,
Ivlev2000}, we have established that 1:1 symmetric and
non-symmetric multibreathers should be expected, while
multibreathers having other resonances are expected to be unstable
and thus physically meaningless.

The methodology which was used to extract these results is quite
general and is directly applicable to a wide variety of systems.
It was shown that the relative strength of electrostatic
inter-particle coupling as compared to the substrate potential
harmonicity, here expressed by the parameter $\epsilon$, may play
a (de-)stabilizing role vis-a-vis discrete breather excitations.
In discharge experiments, the parameter $\epsilon$ may be tuned by
adapting the coupling (e.g., via the grain surface potential,
which determines the charge state $Z$) and the sheath
electrostatic potential (by modifying plasma parameters such as
density or pressure). Therefore, the results of our study may
straightforward be tested (and will hopefully be confirmed) by
appropriate experiments.

From a purely fundamental point of view, one of the principal aims
of this work was to stress the fact that Debye lattices (today
realized as dust-lattices in simple experimental devices) provide
a flexible test-field, available at a convenient meso-/macroscopic
level, for various theoretical and computational predictions now
available concerning oscillatory motion in discrete systems.

Further extensions of our study should take into account
dissipation (here neglected) and/or higher dimensionality effects
(e.g. triangular two-dimensional configurations, as observed in
experiments \cite{Morfill}). Studies in this direction are under
way, and will be reported soon.

\acknowledgments This work was supported by the research programme
PYTHAGORAS II of the Greek Ministry of Education and the E.U. One
of us (I.K.) acknowledges funding from the FWO (Fonds
Wetenschappelijk Onderzoek-Vlaanderen, Flemish Research Fund)
during a visiting appointment at the Sterrenkundig Observatorium,
University of Gent (Belgium). He also acknowledges support from
the Deutsche Forschungsgemeinschaft (DFG, Germany) under the
Emmy-Noether program (grant SH 93/3-1), during the latter stages
of this work. Finally, the authors would like to thank the Max-Planck institute for the Physics of Complex Systems in Dresden and especially S. Flach for accomodating the last part of this work, in March of 2007, during the ``Nonlinear Physics in Periodic Structures and Metamaterials" conference.

\end{document}